\documentclass[prd,aps,showpacs,preprintnumbers,amsmath, amssymb]{revtex4-2}

\usepackage{graphics,epsfig,hyperref}
\usepackage{graphicx}
\usepackage{dcolumn}
\usepackage{bm}
\usepackage{epstopdf}
\usepackage{color}
\usepackage{xcolor}
\usepackage{subfigure}
\usepackage{diagbox} 
\usepackage{adjustbox}
\usepackage{multirow}
\usepackage{amsmath,amssymb,amsfonts}

\usepackage[normalem]{ulem}
\usepackage{xcolor}
\usepackage{mathtools} 
\usepackage{amsthm,amssymb}
\usepackage{mathrsfs}

\begin{document}
	\title{\boldmath Scalar perturbation around a rotating Kalb-Ramond BTZ black hole}

		\author{Zhong-Wu Xia$^{1}$\footnote{zwxia@hunnu.edu.cn}, Sheng Long$^{2}$\footnote{shenglong@ucas.ac.cn}, Huajie Gong$^{1}$\footnote{huajiegong@hunnu.edu.cn}, Qiyuan Pan$^{1,3}$\footnote{panqiyuan@hunnu.edu.cn} and
		Jiliang Jing$^{1,3}$\footnote{jljing@hunnu.edu.cn}}
	\affiliation{	$^1$Department of Physics, Institute of Interdisciplinary Studies, Key Laboratory of Low Dimensional Quantum Structures and Quantum Control of Ministry of Education, Synergetic Innovation Center for Quantum Effects and Applications, and Hunan Research Center of the Basic Discipline for Quantum Effects and Quantum Technologies, Hunan Normal University,  Changsha, Hunan 410081, People's Republic of China} 
	\affiliation{$^2$School of Fundamental Physics and Mathematical Sciences, Hangzhou Institute for Advanced Study, UCAS, Hangzhou 310024, People's Republic of China}
	\affiliation{$^{3}$Center for Gravitation and Cosmology, College of Physical Science and Technology, Yangzhou University, Yangzhou 225009, People's Republic of China
	}
	
	\begin{abstract}
	\baselineskip=0.6 cm
	\begin{center}
		{\bf Abstract}
	\end{center}	
We investigate the scalar perturbation of a newly proposed Kalb--Ramond (KR) BTZ-like black hole. After the separation of variables for the Klein--Gordon equation, we find that the radial part reduces to the general Heun equation. Using the Heun function, we compute quasinormal modes (QNMs) subject to generic Robin boundary conditions, which shows that the KR parameter substantially modifies the QNM spectrum and only the fundamental mode on the left branch has an instability. To ascertain whether the instability is superradiant, we further analyze how the KR field changes the energy and angular momentum fluxes. Our results show that the KR parameter shifts the threshold and the range  of the Robin coupling parameter where the superradiance occurs, underscoring the importance of the KR field in modeling black hole perturbations.

\end{abstract}

	\maketitle
	\flushbottom

	\newcommand{\Heun}{\text{$H_G$}}
	
	\section{Introduction}
	
	Since the first detection of gravitational waves from binary black hole mergers \cite{LIGOScientific:2016aoc}, the LIGO-Virgo-KAGRA collaboration \cite{LIGOScientific:2014pky,VIRGO:2014yos,KAGRA:2020tym} and independent analyses have identified approximately 100 compact binary black hole mergers \cite{LIGOScientific:2017vwq,KAGRA:2021vkt,Olsen:2022pin,Wadekar:2023gea}. These observations have started to reveal the distribution of binary black hole masses and spins \cite{KAGRA:2021duu}, refined the constraints on the neutron star equation of state \cite{LIGOScientific:2018cki,Yang:2022ees}, provided the independent measurements of the Hubble parameter \cite{LIGOScientific:2017adf,LIGOScientific:2021aug}, and validated the general relativity (GR) \cite{LIGOScientific:2016lio,LIGOScientific:2020tif,LIGOScientific:2021sio}.
	
	In the final stage of binary black hole mergers, the black hole generates characteristic oscillations which gradually decays and stabilizes through the radiation of gravitational waves. Therefore, one can describe it by a complex frequency as $\omega = \omega_R + i \omega_I $. Such damped oscillation modes \cite{Kokkotas:1999bd} are called quasinormal modes (QNMs), which directly encode the basic parameters of the black hole, such as mass, charge, angular momentum and  modified gravity parameter, thereby serving as a powerful tool for probing black hole properties \cite{Gogoi:2023kjt,Lan:2023vaa,He:2024drw,Quan:2025tgz,Berti:2025hly,Lo:2025njp,Ma:2024qcv,Deng:2025uvp,Berti:2025hly}. In the gravitational wave physics, the core role of QNMs lies in: through the analysis of frequencies, one can precisely extract black hole parameters, test the validity of general relativity under extreme conditions \cite{Konoplya:2011qq,Nasipak:2019hxh,Mitman:2025hgy}, and explore the potential new physics, such as the quantum gravity~\cite{Gong:2023ghh,Zhang:2024svj,Xia:2023zlf,Skvortsova:2024msa,Konoplya:2025hgp} or dark matter~\cite{Pezzella:2024tkf,Lan:2025brn}.

	QNMs do more than encode the ringdown spectrum, they also diagnose the linear stability of the background spacetime~\cite{Konoplya:2011qq}. When a mode is unstable, i.e., its amplitude grows rather than decays under the standard time dependence, the growth often signals a superradiant mechanism: low-frequency bosonic waves are amplified by extracting rotational energy from the black hole~\cite{Brito:2015oca,Yang:2022uze,Redondo-Yuste:2025ktt}. Such amplification can seed long-lived “scalar clouds” outside the horizon~\cite{Hod:2012px,Benone:2014ssa,Guo:2024bkw}, which is a target that the gravitational wave observations are intended to probe through the precision gravitational wave spectroscopy~\cite{Zhang:2019eid,Dyson:2025dlj,Li:2025ffh}, with direct implications for models of ultralight dark matter. If the amplification is sufficiently strong, the scalar’s backreaction on the geometry becomes non-negligible, naturally leading to “hairy” black hole solutions~\cite{Herdeiro:2014goa,Zhang:2024bfu}. However, recent analyses in the BTZ  spacetime caution against equating all scalar instabilities with the superradiance: due to boundary conditions, the growth can instead originate from the AdS bulk instability~\cite{Ferreira:2017cta,Quan:2025tgz}, demonstrating that the superradiance is sufficient but not necessary for the instability in this setting.
		
	Motivated by the above considerations, we focus on the QNMs and superradiance of a BTZ black hole in the presence of a background Kalb--Ramond (KR) field~\cite{Kalb:1974yc}, as described by the metric provided in Ref. \cite{Liu:2025fxj}. The KR field, a second-rank antisymmetric tensor field associated with the low-energy effective action of superstring theories, introduces a nonminimal coupling to the gravity \cite{Lessa:2019bgi,Yang:2023wtu}. This coupling leads to the modifications in the spacetime geometry, altering the standard Schwarzschild metric and the associated perturbation equations. The framework is particularly motivated by the theories of spontaneous Lorentz symmetry breaking, where the vacuum expectation value (VEV) of the KR field breaks the Lorentz invariance~\cite{Kostelecky2004}. Investigating QNMs in this context is essential for several reasons: First, it allows us to quantify how the Lorentz-violating effects manifest in black hole perturbations, which could have observable implications in the gravitational wave data. Second, such studies bridge the gap between the modified gravity theories and the observational astrophysics, offering a pathway to constrain the parameters of beyond-GR models. Finally, the rotating KR BTZ black hole solutions provide the exact analytic metrics, enabling precise computations that can serve as the benchmarks for more complex scenarios, such as  charged black holes in similar frameworks.

It should be noted that, after separating the Klein--Gordon equation, the radial part is reduced to a Fuchsian equation with four regular singular points, which, by a fractional linear transformation, can be brought into the form of the general Heun equation, extensively discussed in~\cite{olver2010nist}. The determination of QNMs crucially depends on the choice of boundary conditions. For asymptotically flat spacetimes, it is appropriate to impose purely ingoing waves at the event horizon and purely outgoing waves at infinity~\cite{Anabalon:2019zae}. In contrast, for black hole perturbations in the asymptotically anti-de Sitter (AdS) spacetimes, the effective potential does not vanish at the spatial infinity, so the usual outgoing-wave condition cannot be imposed~\cite{Birmingham:1997rj}. In this case, the radial wave function exhibits a power-law decay, and more importantly, both independent solutions corresponding to Dirichlet and Neumann boundary conditions satisfy the requirement of the vanishing energy flux at infinity~\cite{Dappiaggi:2017pbe}. As a consequence, more general mixed boundary,  Robin boundary conditions, become admissible~\cite{Kinoshita:2023iad}. Robin boundary conditions impose a linear relation between Dirichlet and Neumann conditions, and allow for the emergence of additional physical phenomena~\cite{Wang:2015goa,Chen:2024pys,Chen:2023cjd}. By imposing Robin boundary conditions, we determine the QNM spectrum via the connection formula of the Heun equation. Compared with the recently proposed high-precision QNM algorithm based on the Borel summation \cite{Hatsuda:2019eoj}, the Heun equation approach exhibits the agreement up to at least 50 digits~\cite{Hatsuda:2020sbn}. Moreover, owing to the arbitrary-precision computational power of \textit{Mathematica}~\cite{wolfram2003mathematica}, this method can, in principle, yield QNMs with arbitrarily high precision.

	This work is organized as follows. In Section II, we review the rotating KR BTZ black hole and derive the scalar perturbation equation and its reduction to the Heun form. In Section III, we present the results of QNMs and superradiance. In Section IV, we give the conclusion and future direction. We will, in the Appendix, discuss the general Heun Functions in more detail.

\section{Scalar perturbation in a rotating KR BTZ black hole background}

We consider the Einstein--Hilbert gravity nonminimally coupled to a rank--2 antisymmetric KR field \(B_{\mu\nu}\):
	\begin{equation}
		\label{eq:action}
		S=\int d^D x \,\sqrt{-g}\,
		\left[
		\frac{1}{2\kappa}\bigl(R-2\Lambda\bigr)
		-\frac{1}{12} H_{\mu\nu\rho}H^{\mu\nu\rho}
		- V\!\bigl(B_{\mu\nu}B^{\mu\nu}\pm b^2\bigr)
		+\frac{1}{2\kappa}\Bigl(\xi_1 B_{\mu\nu}B^{\mu\nu} R
		+ \xi_2 B_{\rho\mu} B^{\ \mu}_{\nu} R^{\rho\nu}\Bigr)
		\right],
	\end{equation}
with the KR field strength $H_{\mu\nu\rho}\equiv\partial_{[\mu}B_{\nu\rho]}$, the cosmological constant \(\Lambda\) and the gravitational coupling \(\kappa\). Here \(\xi_1\) and \(\xi_2\) denote the nonminimal coupling constants between the KR field and curvature.  The potential \(V\) is chosen so that its minimum enforces a nontrivial VEV for the KR field, and the nonzero VEV selects a specific direction,	leading to the violation of local Lorentz invariance~\cite{Ding:2023niy,Liu:2024oas}.
	
For three-dimensional stationary spacetime ($D=3$), in direct analogy with the four-dimensional, spherically symmetric KR black hole~\cite{Yang:2023wtu}, we adopt a purely ``pseudo-electric'' configuration for the KR field.  Varying the action \eqref{eq:action}  yields a modified Einstein equations, and the corresponding KR BTZ solution is given by~\cite{Liu:2025fxj}
	
	\begin{equation}
		\mathrm{d} s^{2} = - A(\tilde{r}) \mathrm{d} t^{2} + S(\tilde{r}) \mathrm{d} \tilde{r}^{2} + \tilde{r}^{2} [K(\tilde{r}) \mathrm{d} t + \mathrm{d} \varphi]^{2} ,
	\end{equation}
	where
	\begin{align}
		A(\tilde{r}) = & - M - \frac{\Lambda}{(1 + \ell) } \tilde{r}^{2} + \frac{j^{2}}{4 (1 + \ell) \tilde{r}^{2}} , \notag\\
		S(\tilde{r}) = & \frac{1}{A(\tilde{r})} , \notag\\
		K(\tilde{r}) = & - \frac{j}{2 \tilde{r}^{2}} .
	\end{align}
The quantities $M$ and $j$ are  the mass and the spin parameters of the black hole, respectively. $\ell$ is the KR parameter. Here, the cosmological constant $\Lambda=-1/\lambda^2$ with the AdS radius $\lambda$. For simplicity, in the following we rescale the radial coordinate
\begin{equation}
\tilde{r} \rightarrow r = \tilde{r}^{2},
\end{equation}
then obtain the inner and outer horizons	
	\begin{equation}
		r_{\pm} = \frac{1}{2} \lambda \left[\lambda(\ell+1) M\pm\sqrt{\lambda ^2 (\ell+1)^2 M^2-j^2}\right].
	\end{equation}
	
The Klein-Gordon equation in the curved space is \cite{Dappiaggi:2017pbe}
	\begin{align}
		\left( \Box - \frac{\mu^{2}}{\lambda^2} \right) \Phi = 0,
	\end{align}
where $\mu^2/\lambda^2$ is the effective squared mass. For the rotating KR BTZ black hole, we can separate the variables of the scalar field $\Phi(t,r,\varphi) = e^{- i \omega t + i m \varphi} \phi(r)$, so the radial equation is 
	
	\begin{align}
		\nonumber & \left\{ \frac{\mathrm{d}^{2} }{\mathrm{d} r^{2}} + \left( \frac{1}{r - r_{+}} + \frac{1}{r - r_{-}} \right) \frac{\mathrm{d}}{\mathrm{d} r} + \frac{\lambda^{4}(1 + \ell)^{2}}{16 r (r - r_{+})^{2}(r - r_{-})^{2}} \right. \\
		& \left. \times\left[(2 \omega r - m j) ^{2} - \frac{4 (r - r_{-}) (r - r_{+}) (\mu^{2} r + m^{2}\lambda^2)}{\lambda^{4} (1 + \ell)} \right] \right\} \phi(r) = 0 . \label{eq:RadialEqu}
	\end{align}
Applying the coordinate transformation $z = \frac{r - r_{+}}{r - r_{-}}$, we find the point-to-point mapping
	\begin{equation}
		f: r \rightarrow z \Rightarrow (r_{+}, \infty, 0, r_{-}) \rightarrow (0, 1, a, \infty),
	\end{equation}
with $a=r_{+}/r_{-}$. Thus, Eq. \eqref{eq:RadialEqu} becomes
	\begin{equation}
		z (1 - z) {\phi}''(z) + (1 - z) {\phi}'(z) + \left( \frac{A_{0}}{z} + \frac{A_{1}}{z - 1} + \frac{A_{a}}{z - a} + B \right) \phi(z) = 0 ,\label{eq:Heun}
	\end{equation}
	where
	\begin{align}
		A_{0} = & \frac{(2 \omega r_{+} - m j)^{2} (1 + \ell)^{2}}{16 r_{+} (r_{+} - r_{-})^{2}} \lambda^{4} , \notag\\
		A_{1} = & \frac{(1 + \ell) \mu^{2}}{4}\lambda^{2} , \notag\\
		A_{a} = & - \frac{m^{2} j^{2} \ell (1 + \ell)}{16 r_{+} r_{-}^{2}}\lambda^{4} , \notag\\
		B = & - \frac{(2 \omega r_{-} - m j)^{2} (1 + \ell)^{2}}{16 r_{-} (r_{+} - r_{-})^{2} }\lambda^{4} .
	\end{align}
In Eq.~\eqref{eq:Heun}, one has $A_a=0$ when the KR parameter $\ell=0$; in this case, the equation reduces to the standard Gauss hypergeometric equation. With the transformation $\phi(z) \rightarrow z^{\alpha_{1}} (z - 1)^{\beta_{1}} (z - a)^{\delta_{1}} \tilde{\phi}(z)$, we can rewrite Eq.~\eqref{eq:Heun} as
\begin{equation}
{\tilde{\phi}}''(z) + \left( \frac{1 + 2 \alpha_{1}}{z} + \frac{2 \beta_{1}}{z - 1} + \frac{2 \delta_{1}}{z - a} \right) {\tilde{\phi}}'(z) + \sum_{k = 1}^{2} \left[\frac{\Theta_{k}}{z^{k}} + \frac{\Psi_{k}}{(z - 1)^{k}} + \frac{\Delta_{k}}{(z - a)^{k}} \right] \tilde{\phi}(z) = 0 , \label{eq:NewRadialEqu}
\end{equation}
with
	\begin{align}
		\Theta_{1} = & -\frac{a (-A_{0}+2 \alpha_{1} \beta_{1}+A_{1}-B+\beta_{1})+2 \alpha_{1} \delta_{1}+A_{a}+\delta_{1}}{a} , \notag\\
		\Psi_{1} = & \frac{(a-1) (A_{1} + 2 \alpha_{1} \beta_{1} - B + \beta_{1} -A_{0}) + A_{a} - 2 \beta_{1} \delta_{1}}{a-1} , \notag\\
		\Delta_{1} = & \frac{(a-1) (2 \alpha_{1}+1) \delta_{1}+2 a \beta_{1} \delta_{1}-A_{a}}{(a-1) a}, \notag\\
		\Theta_{2} = & A_{0}+\alpha_{1}^2 , \notag\\
		\Psi_{2} = & -A_{1}+\beta_{1}^2-\beta_{1} , \notag\\
		\Delta_{2} = & \delta_{1}^2-\delta_{1}.
	\end{align}
In order to transform Eq. \eqref{eq:NewRadialEqu} into the form of Eq. \eqref{eq:Heun-Equation}, 
we have		
		\begin{equation}
			\begin{cases}
				\alpha = \alpha_{1}+\beta_{1}+\delta_{1} \pm \sqrt{B}, \\
				\beta = \alpha_{1}+\beta_{1}+\delta_{1} \mp \sqrt{B}, \\
				q = a (-A_{0}+2 \alpha_{1} \beta_{1}+A_{1}-B+\beta_{1})+2 \alpha_{1} \delta_{1}+A_{a}+\delta_{1}.
			\end{cases}
		\end{equation}
In our calculation, we will choose 
	\begin{align}
		& (\delta_{1}, \alpha_{1}, \beta_{1}) = \left(0, - i \sqrt{A_{0}}, \frac{1}{2} (1 - \sqrt{1 + 4 A_{1}}) \right), \notag\\
		& (\gamma, \delta, \epsilon) = (1 + 2 \alpha_{1}, 2 \beta_{1}, 0),\notag\\
		& (\alpha, \beta, q) = ( \alpha_{1} + \beta_{1} + \sqrt{B}, \alpha_{1} + \beta_{1} - \sqrt{B}, \alpha \beta a + A_{a}),
	\end{align}
and the corresponding Heun function is $\Heun(a,q;\alpha, \beta, \gamma, \delta; z)$. Because the physical boundary is in the interval $[r_{+}, \infty)$, the solutions we are concerned with to Eq. \eqref{eq:RadialEqu} are
	\begin{align}
		\phi^{(\mathcal I)}(z) = & h(r) u_{01}(z) , \\
		\phi^{(\mathcal O)}(z) = & h(r) u_{02}(z) , \\
		\phi^{(\mathcal N)}(z) = & h(r) u_{11}(z) ,\label{eq:neu} \\ 
		\phi^{(\mathcal D)}(z) = & h(r) u_{12}(z) ,\label{eq:Diri}
	\end{align}
	where $h(r)={(r - r_{+})^{\alpha_{1}} (r_{-} - r_{+})^{\beta_{1}} (1 - a)^{\delta_{1}} r^{\delta_{1}}}/{(r - r_{-})^{\alpha_{1} + \beta_{1} + \delta_{1}}}$. The functions \(u_{01},\,u_{02},\,u_{11}\) and \(u_{12}\) are Heun functions defined in Eqs.~(\ref{eq:u01})--(\ref{eq:u12}). Here \(\mathcal I\) and \(\mathcal O\) denote the ingoing wave and the outgoing wave, and \(\mathcal N\) and \(\mathcal D\) describe the Neumann boundary condition and the Dirichlet boundary condition, respectively.

\section{Heun method and result}
\subsection{Quasinormal modes}
Unfortunately, the local solutions of the Heun equation around distinct regular singular points are not related by the known closed-form connection formula. As a result, unlike the non-KR rotating BTZ spacetime where analytical QNMs can be obtained, here we are unable to obtain analytical QNM spectrum subject to Neumann or Dirichlet boundary conditions \cite{Dappiaggi:2017pbe}. In order to compute the QNM spectrum for the KR rotating BTZ spacetime, we have to make use of hypergeometric functions. Setting the KR parameter $\ell=0$ causes the general Heun equation to degenerate into the Gauss hypergeometric function  ${}_2F_1(z)$. Employing the analytical connection formula for ${}_2F_1$, e.g., Kummer's transformations ~\cite{olver2010nist} which relate local solutions at $z=0,1$, one can directly derive the QNMs of non--KR BTZ black hole under Dirichlet boundary conditions \cite{Dappiaggi:2017pbe}
	\begin{align}
		\omega_{n}^{(\mathcal{D}),\mathrm{L}} &= \frac{m}{\lambda} - i \frac{(\sqrt{r_{+} }- \sqrt{r_{-}})}{\lambda^2} \left(2n + 1 + \sqrt{1 + \mu^2}\right), \label{eq:QNMDL} \\
		\omega_{n}^{(\mathcal{D}),\mathrm{R}} &= -\frac{m}{\lambda} - i \frac{(\sqrt{r_{+} }+ \sqrt{r_{-}})}{\lambda^2} \left(2n + 1 + \sqrt{1 + \mu^2}\right), \label{eq:QNMDR}
	\end{align}
where $n$ denotes the overtone index ($n=0$ is the fundamental mode), and \(L\), \(R\) label the left and right branches, respectively. Accordingly, the QNM frequencies under Neumann boundary conditions are
	\begin{align}
		\omega_{n}^{(\mathcal{N}),\mathrm{L}} &= \frac{m}{\lambda} - i \frac{(\sqrt{r_{+} }- \sqrt{r_{-}})}{\lambda^2} \left(2n + 1 - \sqrt{1 + \mu^2}\right), \label{eq:QNMNL} \\
		\omega_{n}^{(\mathcal{N}),\mathrm{R}} &= -\frac{m}{\lambda} - i \frac{(\sqrt{r_{+} }+ \sqrt{r_{-}})}{\lambda^2} \left(2n + 1 - \sqrt{1 + \mu^2}\right). \label{eq:QNMNR}
	\end{align}
For AdS black holes, both the Neumann boundary condition \eqref{eq:neu} and the Dirichlet boundary condition \eqref{eq:Diri} enforce the vanishing energy flux at inffnity. It is therefore natural to consider a mixed (Robin type)  vanishing energy flux boundary condition~\cite{Ferreira:2017cta}
	\begin{equation}
		\phi^{(\mathcal R)}(z)=\sin(\xi)\phi^{(\mathcal N)}(z)+\cos(\xi) \phi^{(\mathcal D)}(z),~~\xi\in\left[0,\pi\right).
	\end{equation}
In the limits $\xi=0$ and $\xi=\pi/2$, the Robin condition reduces to the Dirichlet boundary condition and Neumann boundary condition, respectively. In order to get the QNMs of the KR BTZ black hole, we employ an ingoing wave boundary condition at the event horizon and the Robin boundary condition at infinity. Thus, the QNMs are determined by the vanishing of the Wronskian
\begin{equation}\label{eq:wronskian_def}
	W(z)=\phi^{(\mathcal R)}(z)\frac{\mathrm d}{\mathrm d z}\phi^{(\mathcal I)}(z)-\phi^{(\mathcal I)}(z)\frac{\mathrm d}{\mathrm d z}\phi^{(\mathcal R)}(z)=0,
\end{equation}
where $\phi^{(\mathcal I)}$ represents the ingoing wave. In our calculation, the QNM condition is $W\!\left(1/2\right)=0$ for simplicity.

\begin{figure}[tbp]
	\centering
	\begin{minipage}[b]{0.32\textwidth}
		\centering
		\includegraphics[width=\textwidth]{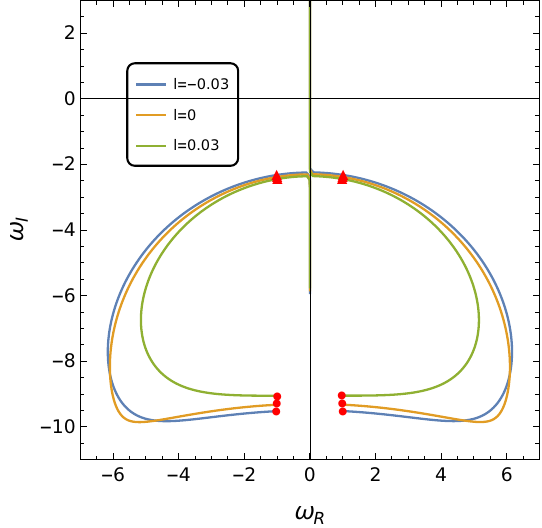}
	\end{minipage}
	\begin{minipage}[b]{0.32\textwidth}
		\centering
		\includegraphics[width=\textwidth]{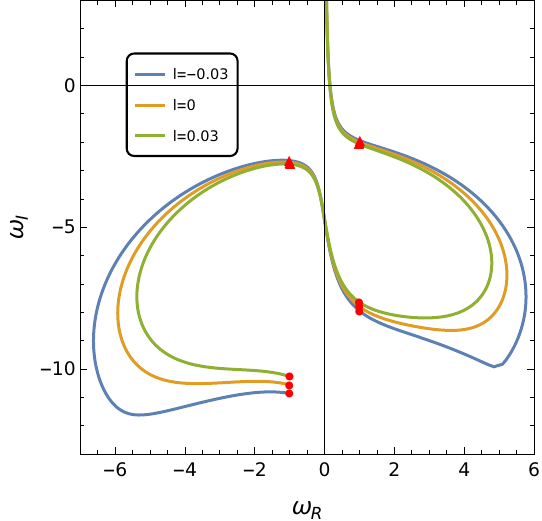}
	\end{minipage}
	\begin{minipage}[b]{0.32\textwidth}
		\centering
		\includegraphics[width=\textwidth]{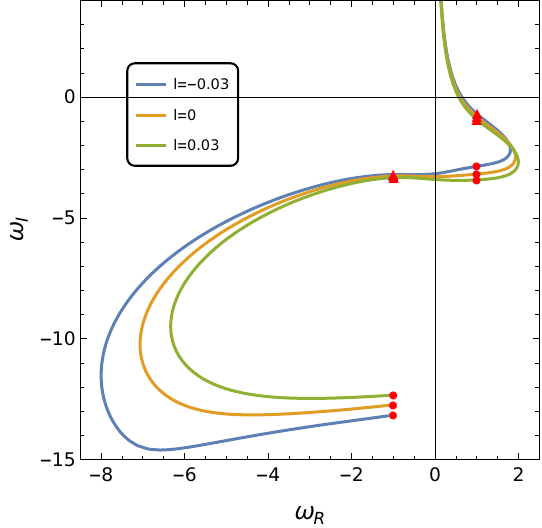}
	\end{minipage}
	
	\caption{Trajectories in the real part $\omega_R$ and imaginary parts $\omega_I$  of selected QNM frequencies with $n=0$ as the coupling $\xi$ varies for a KR BTZ black hole with different KR parameters $\ell=-0.03$, $0$, $0.03$ with $\mu^2=-0.65$, $m=1$, $\lambda=1$ and $M=34$. Panels from left to right correspond to angular momenta $j=0, 10$, and $30$. Filled circles and triangles denote solutions satisfying Dirichlet and Neumann boundary conditions, respectively. 
	}
	\label{fig:Re_Im_QNM}
\end{figure}

\begin{figure}[tbp]
	\centering
	\begin{minipage}[b]{0.32\textwidth}
		\centering
		\includegraphics[width=\textwidth]{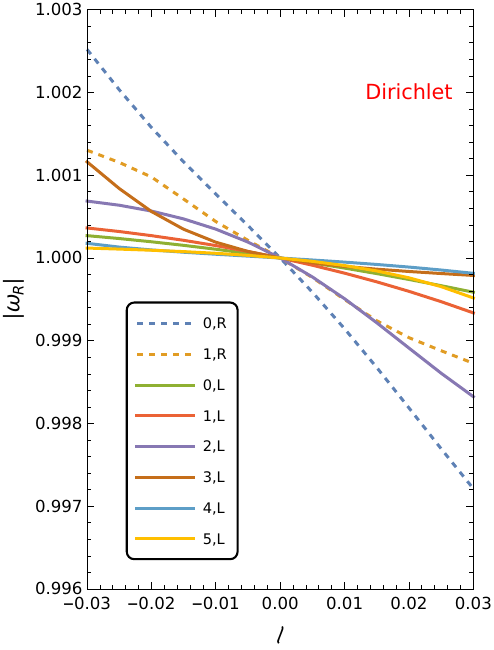}
	\end{minipage}
	\begin{minipage}[b]{0.31\textwidth}
		\centering
		\includegraphics[width=\textwidth]{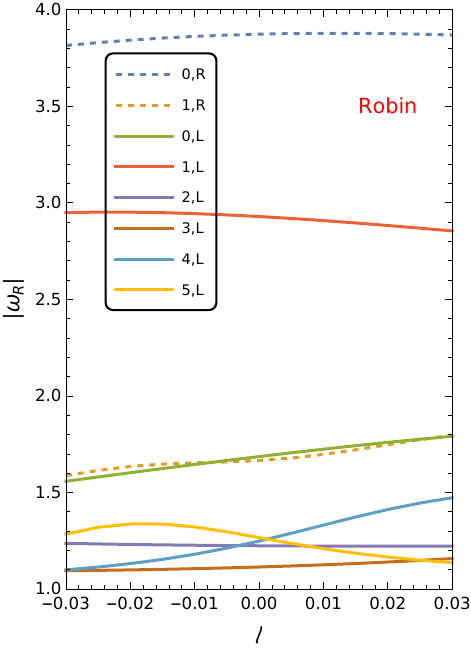}
	\end{minipage}
	\begin{minipage}[b]{0.32\textwidth}
		\centering
		\includegraphics[width=\textwidth]{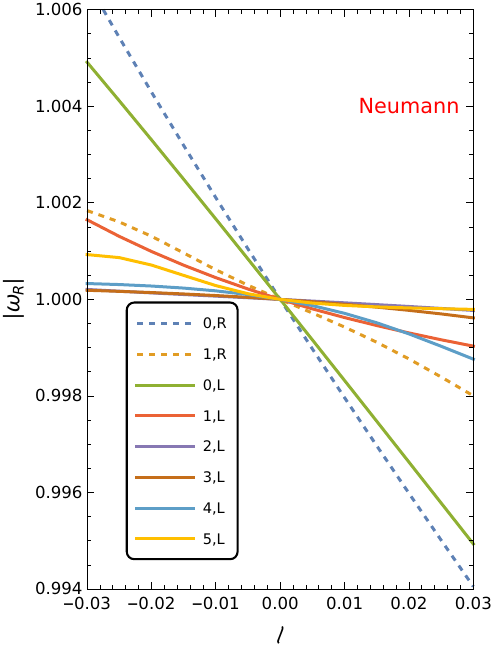}
	\end{minipage}
	\begin{minipage}[b]{0.32\textwidth}
		\centering
		\includegraphics[width=\textwidth]{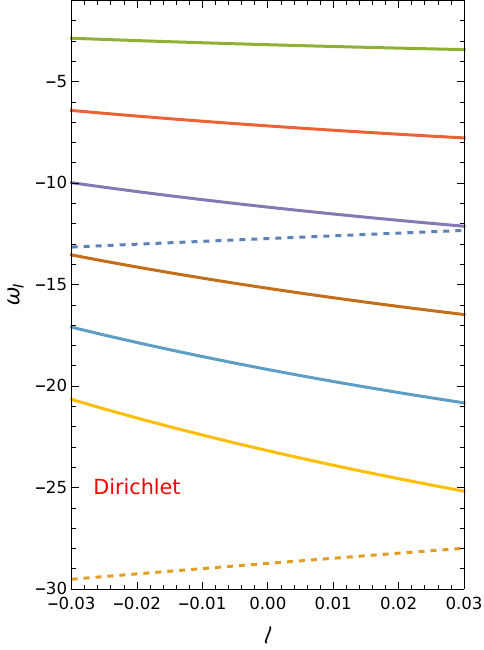}
	\end{minipage}
	\begin{minipage}[b]{0.32\textwidth}
		\centering
		\includegraphics[width=\textwidth]{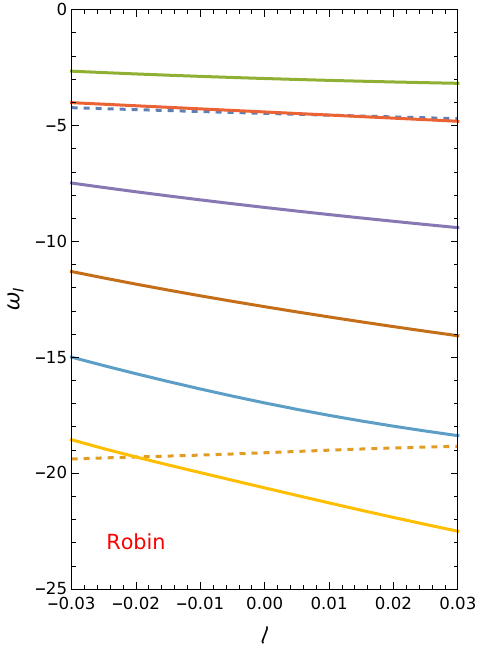}
	\end{minipage}	
	\begin{minipage}[b]{0.32\textwidth}
		\centering
		\includegraphics[width=\textwidth]{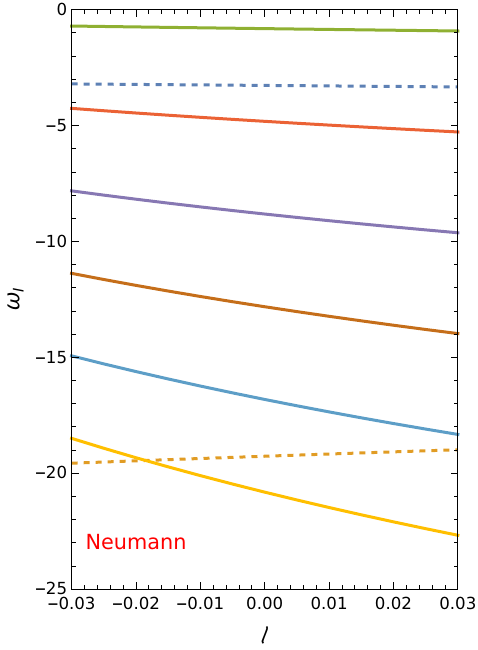}
	\end{minipage}	
	\caption{
		Real and imaginary parts of some QNM frequencies as a function of KR parameter $\ell$ with $\mu^2=-0.65$, $m=1$, $\lambda=1$,  $M=34$ and $j=30$.
		The left two panels display the real and imaginary parts of the QNMs for Dirichlet boundary conditions ($\xi=0$), the middle panels for Robin boundary conditions ($\xi=\pi/4$), and the right panels for Neumann boundary conditions ($\xi=\pi/2$).	
	}
	\label{fig:DRN_QNM}
\end{figure}	

In calculation, we denote the step size of the iterative procedure by $\Delta\xi$ and $\Delta\ell$. For the non--KR BTZ black hole ($\ell=0$), provided $\Delta\xi$ is sufficiently small, any known quasinormal frequency $\omega(\xi)$ can be continued to $\omega(\xi+\Delta\xi)$ by numerically solving the Wronskian~\eqref{eq:wronskian_def} using $\omega(\xi)$ as the initial guess.
To accelerate the computations, we employed \textit{Mathematica}'s multi-process capabilities \cite{wolfram2003mathematica}. For the KR black hole ($\ell\neq0$), we employ the Dirichlet quasinormal frequencies $\omega_n^{(\mathcal D)}$ (see Eqs.~(\ref{eq:QNMDL})--(\ref{eq:QNMDR})) or the Neumann frequencies $\omega_n^{(\mathcal N)}$ (see Eqs.~(\ref{eq:QNMNL})--(\ref{eq:QNMNR})) at $\ell=0$ as initial guesses. Similarly,
we obtain $\omega_n^{(\mathcal D/ \mathcal N)}(\Delta\ell)$ by numerically solving the Wronskian \eqref{eq:wronskian_def} with $\omega_n^{(\mathcal D/ \mathcal N)}(0)$. 
So we can find $\omega_n^{(\mathcal D/ \mathcal N)}$ for arbitrary $\ell$. Then starting from $\omega_n^{(\mathcal D/ \mathcal N)}(\ell)$, one can solve for arbitrary Robin QNMs $\omega_n^{(\mathcal R)}(\ell)$ by continuously varying $\xi$.

In Fig.~\ref{fig:Re_Im_QNM} we present the fundamental QNM spectrum obtained numerically by using general Heun functions. We show the relation between $\omega_R$ and $\omega_I$ as the coupling $\xi$ increases from $0$ to $\pi$, yielding two frequency branches (left and right). In the case of $j=0$, the two branches are exactly symmetric; introducing the KR parameter preserves this symmetry but produces a pronounced shift of the QNM frequencies. As the angular momentum $j$ increases, the symmetry is progressively broken. We further find that the KR parameter substantially modifies the spectrum under all boundary conditions considered. 

To further understand the impact of the KR parameter $\ell$ on quasinormal modes under different boundary conditions, in Fig.~\ref{fig:DRN_QNM} we fix $\xi=0$, $\pi/4$ and $\pi/2$, corresponding to Dirichlet, Robin and Neumann boundary conditions, respectively. For Dirichlet and Neumann conditions,  increasing $\ell$ reduces the real parts of the frequencies though the effect is small.  For the Robin conditions, the dependence on $\ell$ is more subtle: the absolute values of real parts of $\omega_{0}^{(\mathcal R),R}$, $\omega_{1}^{(\mathcal R),R}$, $\omega_{0}^{(\mathcal R),L}$, $\omega_{3}^{(\mathcal R),L}$ and $\omega_{4}^{(\mathcal R),L}$ increase but 
$\omega_{1}^{(\mathcal R),L}$ and $\omega_{2}^{(\mathcal R),L}$
decrease as $\ell$ increases, whereas $\omega_{5}^{(\mathcal R),L}$ first increases and then decreases. For the imaginary parts, most modes exhibit a decrease with $\ell$, which indicates that
the KR parameter makes the QNMs decay faster, except for $\omega_{0}^{(\mathcal D),R}$, $\omega_{1}^{(\mathcal D),R}$, $\omega_{1}^{(\mathcal R),R}$, and $\omega_{1}^{(\mathcal N),R}$.

	\begin{figure}[tbp]
	\centering
	\begin{minipage}[b]{0.32\textwidth}
		\centering
		\includegraphics[width=\textwidth]{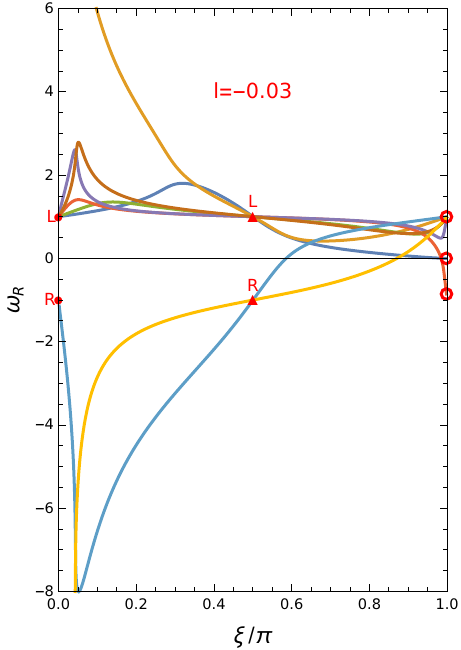}
	\end{minipage}
	\begin{minipage}[b]{0.32\textwidth}
		\centering
		\includegraphics[width=\textwidth]{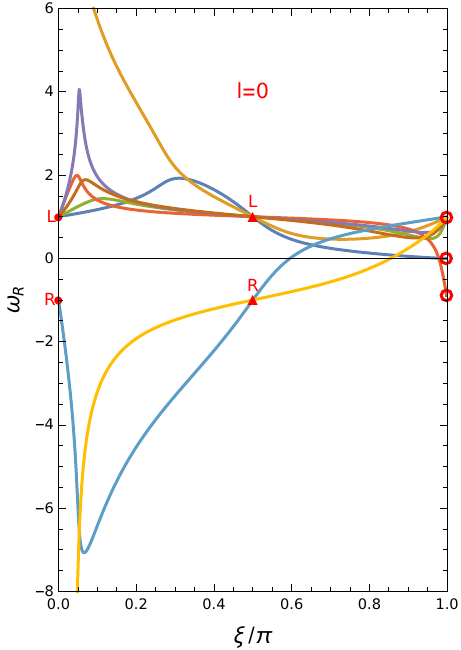}
	\end{minipage}
	\begin{minipage}[b]{0.32\textwidth}
		\centering
		\includegraphics[width=\textwidth]{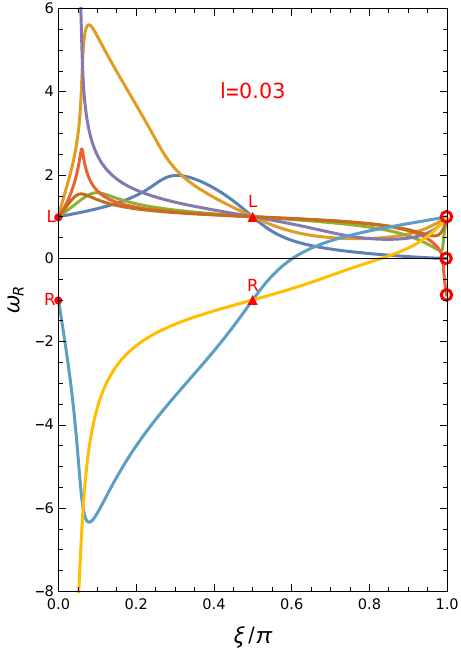}
	\end{minipage}
	\begin{minipage}[b]{0.32\textwidth}
		\centering
		\includegraphics[width=\textwidth]{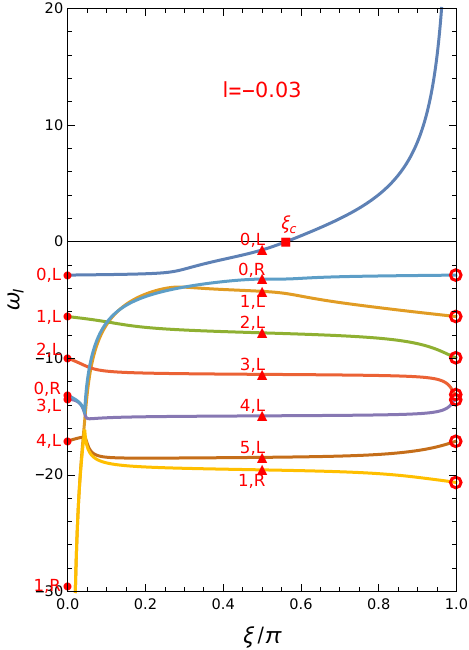}
	\end{minipage}
	\begin{minipage}[b]{0.32\textwidth}
		\centering
		\includegraphics[width=\textwidth]{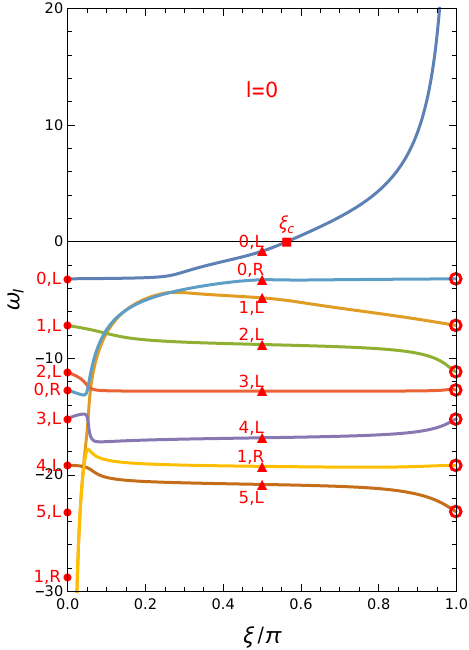}
	\end{minipage}
	\begin{minipage}[b]{0.32\textwidth}
		\centering
		\includegraphics[width=\textwidth]{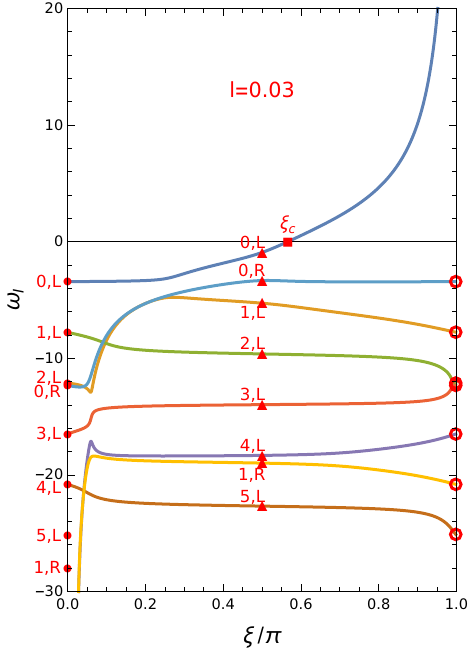}
	\end{minipage}	
	
	\caption{Real and imaginary parts of some QNM frequencies as a function of $\xi/\pi$ for a KR BTZ black hole with different KR parameters $\ell=-0.03$, $0$, $0.03$ with $\mu^2=-0.65$, $m=1$, $\lambda=1$,  $M=34$ and $j=30$. The filled circles and triangles denote QNMs under Dirichlet boundary conditions and Neumann boundary conditions, respectively. The squares represent QNMs with the zero imaginary part, and from left to right the corresponding  thresholds are $\xi_c/\pi=0.561393$, $0.563514$ and $0.566042$. }
	\label{fig:QNM_spectrum}
	
\end{figure}

In Fig.~\ref{fig:QNM_spectrum} we display the QNMs as a function of $\xi/\pi$ for a KR BTZ black hole with different KR parameters $\ell=-0.03$, $0$ and $0.03$. We find that, starting from a QNM under the Neumann boundary condition and continuously decreasing $\xi$, one mode always approaches to a Dirichlet QNM as $\xi \to 0$, but the overtone number $n$ does not always coincide. For example, from the imaginary parts of QNM frequencies in Fig.~\ref{fig:QNM_spectrum}, we have $\omega^{(\mathcal N),L}_{3} \to \omega^{(\mathcal D),L}_{2}$ when $\ell=-0.03$ and $0$, but $\omega^{(\mathcal N),L}_{3} \to \omega^{(\mathcal D),L}_{3}$ when $\ell=0.03$. Thus, it is interesting to note that this overtone number correspondence depends on the value of the KR parameter. Although no simple rule emerges for this overtone number correspondence, the branch identity is preserved as $\xi$ varies: left branch modes remain on the left, and right branch modes remain on the right. Moreover, we observe a hint of superradiance only on the left branch for the QNMs with $n=0$, i.e., the imaginary part of the fundamental QNM frequency 
changes from negative to positive as the parameter $\xi$ increases, and the corresponding threshold $\xi_c$ increases as the KR parameter $\ell$ increases. Interestingly, regardless of the KR parameter $\ell$, the superradiance  only appears in
the fundamental modes ($n=0$) of the left branch.

\subsection{Superradiance}

In black hole physics, the superradiance usually refers to the phenomenon whereby ingoing waves on a rotating black hole need not always be damped — in certain frequency ranges they can extract energy and angular momentum from the hole and be reflected with an amplitude larger than the ingoing one~\cite{Brito:2015oca,Baryakhtar:2017ngi,Zhang:2020sjh,Gonzalez:2021vwp}. This mechanism is intimately tied to the reservoir of rotational energy of the black hole. In AdS spacetimes, the reflective nature of the boundary allows perturbations to bounce back and forth, which can enhance or qualitatively change the superradiance. As pointed out in \cite{Dappiaggi:2017pbe}, the modes with positive imaginary parts ($\omega_I>0$) in an AdS background do not necessarily correspond to the superradiance; they can also originate from a bulk (volume) instability. Hence the condition $\omega_I>0$ alone is insufficient to conclude that the superradiance occurs. To examine whether the KR BTZ spacetime exhibits the superradiant extraction, we must compute the energy and angular momentum fluxes at the event horizon and check whether the energy and angular momentum are being extracted.
	
To facilitate the analysis near the event horizon, we employ ingoing Eddington--Finkelstein (EF) coordinates $(v,r,\hat{\varphi})$. The motivation is to avoid coordinate singularities at the horizon and thereby provide a regular description of ingoing waves. The coordinate transformation is
	\begin{equation}
		dv = dt + dr_* \doteq dt + \frac{dr}{A},\qquad
		d\hat{\varphi} = d\varphi - \frac{K}{A}\,dr,
	\end{equation}
	where the tortoise coordinate $r_*$ is defined by
	\begin{equation}
		r_* = \frac{1+\ell}{2\left(r_+^2-r_-^2\right)}
		\Bigg[r_+ \log\!\Big(\frac{r-r_+}{r+r_+}\Big)
		- r_- \log\!\Big(\frac{r-r_-}{r+r_-}\Big)\Bigg].
	\end{equation}
In this coordinate, we have  $r_* \to -\infty$ as $r\to r_+$, ensuring that EF time slices extend smoothly through the horizon. Near the event horizon, the scalar field solution should be a purely ingoing wave
	\begin{equation}
		\Phi(v,r_+,\hat{\varphi})=\phi^{(I)}(z)\,e^{-i\omega t + i m\varphi}
		= \mathcal A\,e^{-i\omega v + i m \hat{\varphi}},
	\end{equation}
where the normalization constant is
	\begin{equation}
		\mathcal A=\Bigg(\frac{4r_+^2}{r_+^2-r_-^2}\Bigg)^{\!\alpha},
	\end{equation}
with $\alpha$ fixed by the boundary condition. In EF coordinates, this form describes an ingoing mode propagating along the advanced time $v$.

For a massive scalar field $\Phi$, the energy momentum tensor is
	\begin{equation}
		T_{\mu\nu}
		= \partial_{(\mu}\bar{\Phi}\,\partial_{\nu)}\Phi
		- \tfrac{1}{2}\,g_{\mu\nu}\!\left(g^{\rho\lambda}\partial_{(\rho}\bar{\Phi}\,\partial_{\lambda)}\Phi
		+ \tfrac{1}{2}\frac{\mu^2}{\lambda^2}|\Phi|^2\right),
	\end{equation}
which leads to the horizon energy flux
	\begin{equation}
		\mathcal{F}_E(v)
		=\int_0^{2\pi} d\hat{\varphi}\; r_+\,T_{\mu\nu}\,\chi^\mu k^\nu
		= F\big[\omega_I^2+\omega_R(\omega_R-m\Omega_{\mathcal{H}})\big]\,e^{2\omega_{I} v},
	\end{equation}
with the constant $F=2\pi r_+ \mathcal A^2$, the radial basis vector $k^\nu=\partial^\nu_r$, the angular velocity of the event horizon $\Omega_{\mathcal{H}}= \sqrt{r_-}/(\lambda \sqrt{r_+})$ and the horizon Killing vector $\chi^\mu=\partial_t+\Omega_{\mathcal{H}}\partial_{\hat\varphi}$. The two terms inside the brackets represent distinct effects: $\omega_I^2$ captures the time decay/growth, while $\omega_R(\omega_R-k\Omega_{\mathcal{H}})$ encodes the condition for the energy exchange with the rotating black hole. If $\omega_R<m\Omega_{\mathcal{H}}$, the energy flux $\mathcal{F}_E$ can become negative, indicating that the energy is being extracted from the black hole, which is known as the superradiance. Similarly, the angular momentum flux at the horizon is
	\begin{equation}
		\mathcal{F}_L(v)
		= -\int_0^{2\pi} d\hat{\varphi}\; r_+\,T_{\mu\nu}\,\chi^\mu m^\nu
		= F\,m(\omega_R-m\Omega_{\mathcal{H}})\,e^{2\omega_I v},
	\end{equation}
where $m^\nu=\partial^\nu_{\hat\varphi}$ is the axial Killing vector. This expression shows that, when $\omega_R<m\Omega_{\mathcal{H}}$, the angular momentum is extracted together with the energy. By jointly analyzing the signs and temporal evolution of $\mathcal{F}_E$ and $\mathcal{F}_L$, one can unambiguously distinguish the superradiant extraction from bulk instabilities and thereby establish a stability criterion for the KR BTZ spacetime.

	\begin{figure}[tbp]
		\centering
		\begin{minipage}[b]{0.49\textwidth}
			\centering
			\includegraphics[width=\textwidth]{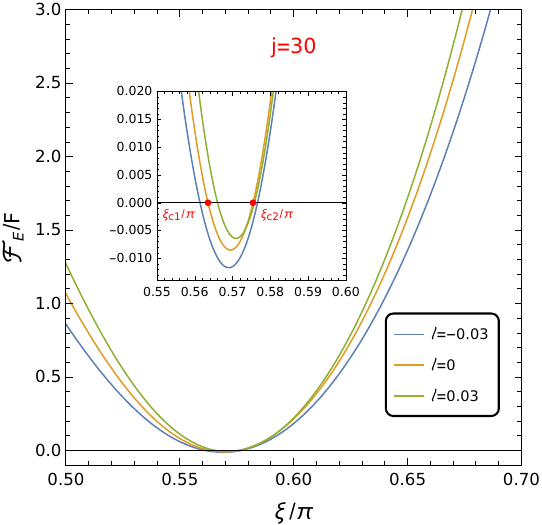}
		\end{minipage}
		\vspace{0.5cm} 
		\begin{minipage}[b]{0.475\textwidth}
			\centering
			\includegraphics[width=\textwidth]{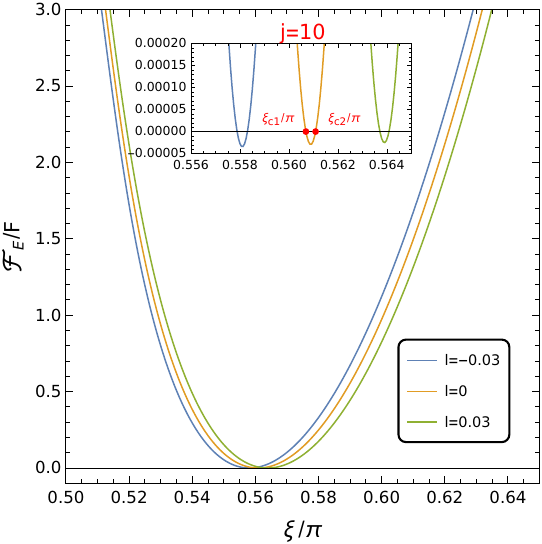}
		\end{minipage}	
		\begin{minipage}[b]{0.48\textwidth}
	\centering
		\includegraphics[width=\textwidth]{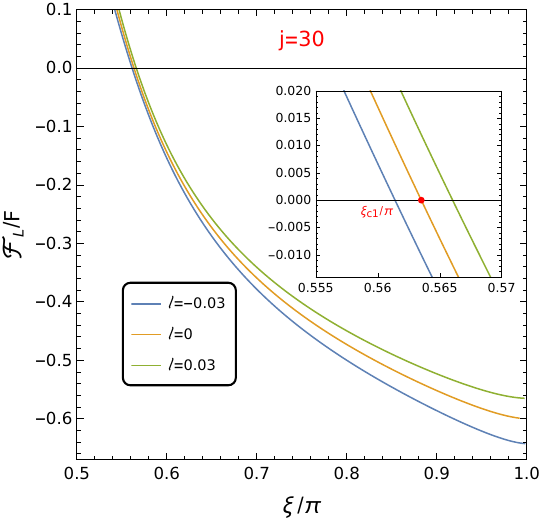}
	\end{minipage}
	\begin{minipage}[b]{0.49\textwidth}
		\centering
		\includegraphics[width=\textwidth]{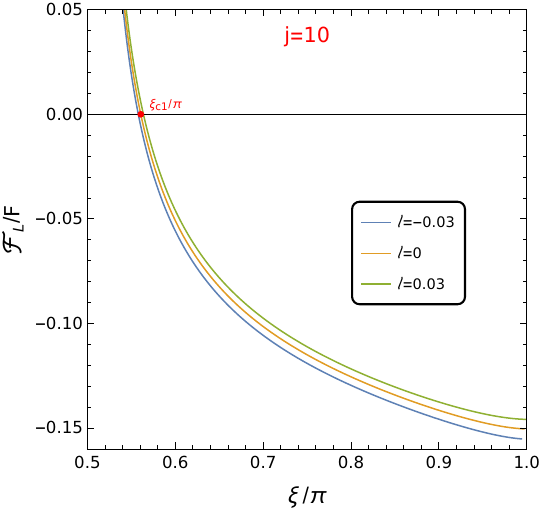}
	\end{minipage}			
	\caption{Energy and angular momentum fluxes as functions of $\xi/\pi$ for a KR BTZ black hole with different KR parameters $\ell=-0.03$, $0$, $0.03$ with $n=0$, $\mu^2=-0.65$, $m=1$, $\lambda=1$ and $M=34$. The left panels are $j=30$ and right panels are $j=10$. Here, the  thresholds of $\ell=0$ for $j=30$ are $\xi_{c1}=0.563514$ and $\xi_{c2}=0.575424$, and the right ones of $\ell=0$ for $j=10$  are $\xi_{c1}=0.560693$ and $\xi_{c2}=0.561089$.}
		\label{fig:fun_sup1}

	\end{figure}

In Fig.~\ref{fig:fun_sup1}, we show the energy and angular momentum fluxes as functions of $\xi/\pi$ for different KR parameters $\ell=-0.03$, $0$, $0.03$ with $j=30$ (left panels) and $j=10$ (right panels). From Fig.~\ref{fig:QNM_spectrum}, we observe that as $\xi$ crosses the threshold $\xi_{c}$, the imaginary part of the quasinormal frequency changes the sign: $ \omega_I<0\to \omega_I>0$, signaling the onset of an instability (the mode grows rather than decays). At the same threshold (for example, $\xi_{c}=\xi_{c1}=0.563514$ when $\ell=0$ presented in Figs.~\ref{fig:QNM_spectrum} and \ref{fig:fun_sup1}), the energy and angular momentum fluxes, $\mathcal{F}_E$ and $\mathcal{F}_L$, flip from positive to negative. With our sign convention, a negative horizon flux corresponds to the extraction from the black hole, so the instability emerging for $\xi>\xi_{c1}$ is superradiant and $\xi_{c1}$ marks the low bound of the superradiance. For a larger $\xi$, one still finds $\omega_I>0$ even if $\mathcal{F}_E$ turns positive again when $\xi>\xi_{c2}$, indicating that the continued growth of $\omega_{I}$ (in Fig.~\ref{fig:QNM_spectrum}) is no longer powered by the horizon extraction but instead by a bulk AdS instability. Turning on the KR deformation ($\ell\neq 0$) preserves this qualitative picture while significantly shifting the thresholds and amplitudes. Within the parameter set of Fig.~\ref{fig:fun_sup1}, increasing $\ell$ moves the superradiant onset $\xi_{c1}$ to a larger value and narrows the parameter range that supports the superradiance. Examining the right panels of $j=10$ confirms the same trends and the extent of superradiance are smaller than that in the case of $j=30$.

	\section{Conclusion}
	\label{sec:conclusion}
	We have analyzed the scalar perturbation of a rotating BTZ-like black hole coupled to a background KR field. Upon the separation of variables, the radial Klein--Gordon equation reduces to the general Heun equation, enabling a direct construction of QNMs under generic Robin boundary conditions. Numerically, we determined the QNM spectrum by matching local Heun solutions via Wronskians, using the analytically tractable $\ell=0$ (non-KR) hypergeometric limit as seeds for the continuation in the KR parameter $\ell$. We reproduced the known BTZ results at $\ell=0$ and observed that the KR parameter $\ell$ substantially modifies the QNM spectrum under all boundary conditions considered. It should be noted that, as the angular momentum $j$ increases, the symmetry of QNMs in the complex plane is progressively broken. We proposed a new overtone number correspondence between the Dirichlet and Neumann boundary conditions, which depends on the value of $\ell$. We further found that the imaginary part of the fundamental QNM  ($n=0$) frequency on the left branch changes from negative to positive as the Robin coupling parameter $\xi$ increases, and the corresponding threshold $\xi_c$ increases as $\ell$ increases, which implies that the superradiance  only appear in the fundamental modes of the left branch. 

	Moreover, a flux-based diagnosis in ingoing EF coordinates shows that the onset of the instability correlates with the horizon energy and angular momentum extraction (corresponding to the superradiance), while the regimes with the mode growth but the energy absorption correspond to the bulk AdS instabilities rather than the superradiance. We observed that the KR parameter $\ell$ strongly modulates  the threshold  of superradiance which just agrees well with  $\xi_c$. We also found that, increasing \(\ell\) reduces the parameter range of $\xi$ where the superradiance occurs, indicating that the positive KR parameter exerts a suppressing effect on superradiance, while a negative KR parameter promotes it. Thus, the KR parameter $\ell$  is a key ingredient in the perturbative stability of BTZ-like geometries. 
	
	A natural next step is to extend the present scalar-field analysis to higher-spin perturbations, including Dirac (spin-$1/2$), electromagnetic (spin-$1$) and gravitational (spin-$2$) fields, to test whether the KR-induced trends identified here still hold for different spin perturbations. Such an extension would provide a more comprehensive assessment of the stability of KR BTZ-like black holes. Furthermore, the presence of superradiance motivates investigating scalar clouds that may form in these spacetimes. More broadly, our flux-based criteria in the AdS spacetime provides a template for disentangling the genuine superradiance from bulk instabilities in related beyond-GR settings.

	\acknowledgments
	
The authors are deeply grateful to Dr. Qin Tan, Chengjia Chen and Fangli Quan for useful discussions. This work was supported by the National Natural Science Foundation of China (Grant Nos. 12275079 and 12035005), the National Key Research and Development Program of China (Grant No. 2020YFC2201400),  the innovative research group of Hunan Province under Grant No. 2024JJ1006
Natural Science Foundation of China (Grants No. 12447156) and China Post-doctoral Science Foundation (Grant No. 2025M773339).

\appendix

\section{General Heun Functions}

A second-order ordinary differential equation in which all singularities are regular is called a Fuchsian equation. One of its most important special cases is the equation with three regular singular points, whose prototype is the hypergeometric equation, while the case with four regular singular points corresponds to the Heun equation~\cite{Heun1888ZurTD,olver2010nist,Hortacsu:2011rr}.

\begin{equation} \label{eq:Heun-Equation}
	{u}''(z) + \left( \frac{\gamma}{z} + \frac{\delta}{z - 1} + \frac{\epsilon}{z - a} \right) {u}'(z) + \frac{\alpha \beta z - q}{z (z - 1) (z - a)} u(z) = 0 .
\end{equation}
The four singular points lie at $z = 0$, $1$, $a$ and $\infty$, where  $\alpha$, $\beta$, $\gamma$, $\delta$ and $\epsilon$ are the exponent parameters satisfying  the  relation $\alpha + \beta + 1 = \gamma + \delta + \epsilon$, and $q$ is the accessory parameter. The Frobenius solution of equation \eqref{eq:Heun-Equation} at the singularity $z =0$ is denoted as $\Heun (a, q;\alpha, \beta, \gamma, \delta; z)$, called the general Heun function, which can be expressed as
\begin{equation}
	\Heun(a, q ; \alpha, \beta, \gamma, \delta ; z)=\sum_{j=0}^{\infty} c_j z^j, \quad|z|<1,
\end{equation}
with
\begin{align}
	& a \gamma c_1-q c_0 = 0 , \\
	R_j c_{j+1}- & \left(Q_j+ q\right) c_j+P_j c_{j-1}=0, \quad j \geq 1,
\end{align}
and
\begin{align}
	P_j & =(j-1+\alpha)(j-1+\beta), \\
	Q_j & =j((j-1+\gamma)(1+a)+a \delta+\epsilon), \\
	R_j & =a(j+1)(j+\gamma),
\end{align}
where $c_0=1$. The Heun function $\Heun(a, q; \alpha, \beta, \gamma, \delta; z)$ is normalized such that
\begin{equation}
	\Heun(a, q; \alpha, \beta, \gamma, \delta; 0) = 1.
\end{equation}

According to the theory of differential equations, the local solutions of Eq.~\eqref{eq:Heun-Equation} around the singularity $z=0$ depend on the value of $\gamma$. When $\gamma \notin \mathbb{Z}$, the two linearly independent Frobenius solutions take the form~\cite{olver2010nist}
\begin{align}
	u_{01}(z) = & \Heun(a,q;\alpha, \beta, \gamma, \delta; z) , \label{eq:u01}
	\\
	u_{02}(z) = & z^{1 - \gamma} \Heun(a,(a \delta + \epsilon)(1 - \gamma) + q;\alpha + 1 - \gamma, \beta + 1 - \gamma, 2 - \gamma, \delta; z) . \label{eq:u02}
\end{align}
Similarly, the two solutions at $z = 1$ are
\begin{align}
	u_{11}(z) = & \Heun(1- a, \alpha \beta - q; \alpha, \beta, \delta, \gamma; 1 - z) ,
	\label{eq:u11}
	\\
	\nonumber
	u_{12}(z) = & (1 - z)^{1 - \delta} \times \\
	& \Heun(1 - a, ((1-a) \gamma + \epsilon)(1 - \delta) + \alpha \beta - q;\alpha + 1 - \delta, \beta + 1 - \delta, 2 - \delta, \gamma; 1 - z) . \label{eq:u12}
\end{align}
There are connection relations between the local solutions in different domains. For instance, one may write 
\begin{align}
	u_{01}(z) = & \mathcal{C}_{11} u_{11}(z) + \mathcal{C}_{12} u_{12}(z) , \\
	u_{02}(z) = & \mathcal{C}_{21} u_{11}(z) + \mathcal{C}_{22} u_{12}(z) .
\end{align}
However, in contrast to hypergeometric functions, no closed-form connection formulas are available for the coefficients $\mathcal{C}_{ij}$, so alternative approaches must be employed. 
The determination of the connection coefficients $\mathcal{C}_{ij}$ is typically carried out by two main methods: (1) the analytical continuation combined with phase restrictions (primarily for connection problems between regular and irregular singular points in confluent Heun functions, see e.g.~\cite{Chen:2025sbz, Bonelli:2022ten, Bonelli:2021uvf}); (2) the numerical evaluation through the Wronskians of local solutions~\cite{Hatsuda:2020sbn}.

	
	\bibliography{reference.bib}
	
	
		
		
		
		
\end{document}